\documentclass[aps, prl, preprint, superscriptaddress, showpacs, amssymb]{revtex4}

\usepackage{graphicx}

\begin{document}

\title{Microscopic Theory of Cation Exchange in CdSe Nanocrystals}

\author{Florian D. Ott} \affiliation{Optical Materials Engineering Laboratory, ETH Zurich, 8092 Zurich, Switzerland}
\author{Leo L. Spiegel}    \affiliation{Optical Materials Engineering Laboratory, ETH Zurich, 8092 Zurich, Switzerland}
\author{David J. Norris}\email{dnorris@ethz.ch}\affiliation{Optical Materials Engineering Laboratory, ETH Zurich, 8092 Zurich, Switzerland}
\author{Steven C. Erwin}\email{steve.erwin@nrl.navy.mil} \affiliation{Center for Computational Materials Science, Naval Research Laboratory, Washington, D.C. 20375, USA}

\date{September 15, 2014}

%%%%%%%%%%%%%%%%%%%%%%%%%%%%%%%%%%%%%%%%%%%%%%%%%%%%%%%%%%%%%%%%%%%%%
%% The manuscript does not need to include \maketitle, which is
%% executed automatically.  The document should begin with an
%% abstract, if appropriate.  If one is given and should not be, the
%% contents will be gobbled.
%%%%%%%%%%%%%%%%%%%%%%%%%%%%%%%%%%%%%%%%%%%%%%%%%%%%%%%%%%%%%%%%%%%%%

\begin{abstract}
  Although poorly understood, cation-exchange reactions are
  increasingly used to dope or transform colloidal semiconductor
  nanocrystals (quantum dots).  We use density-functional theory and
  kinetic Monte Carlo simulations to develop a microscopic theory that
  explains structural, optical, and electronic changes observed
  experimentally in Ag-cation-exchanged CdSe nanocrystals. We find
  that Coulomb interactions, both between ionized impurities and with
  the polarized nanocrystal surface, play a key role in 
  cation exchange.  Our theory also resolves several experimental
  puzzles related to photoluminescence and electrical behavior in
  CdSe nanocrystals doped with Ag.
\end{abstract}

\maketitle

Colloidal nanocrystals \cite{Klimov} with well-defined size and shape
can be synthesized directly only for a few semiconductors. Each new
material requires considerable effort to optimize the growth
\cite{Bawendi}. To avoid this, researchers have exploited a simple
process known as ``cation exchange'' \cite{Son2004, Mokari2006,
Manna2011a, Rivest2013, Brandon2013, Gupta2013}. Nanocrystals (NCs)
that can already be directly synthesized (e.g.~CdSe) are exposed in
solution to cations (e.g.~Ag$^{+}$) that quickly diffuse into the
lattice and replace the original cations.  If this reaction goes to
completion, a new crystalline material is created (e.g.~Ag$_2$Se) with
the size and shape of the initial NCs.  Recently, cation exchange has
also been used for another purpose: to incorporate impurities as
dopants into NCs
\cite{Banin2011, Ayash2012}. Dopants can create mobile
electrons and holes but
their incorporation during the conventional growth of NCs has been
challenging \cite{Norris2008}.  By arresting cation exchange at an
early stage, a controllable number of impurities can be
introduced by diffusion into premade NCs.  Cation exchange is
thus becoming an important route to both new and doped NCs. However,
the mechanisms underlying cation exchange remain unclear.
A microscopic theory could greatly extend the use
of cation-exchange reactions for creating new NC materials.

The theory must explain several experimental observations for the
well-studied cation exchange of Ag in Cd$E$ NCs ($E$ = S or Se). (1) When
sufficient Ag is added to solution, Ag$_2E$ NCs are obtained very
rapidly \cite{Son2004}. (2) Early attempts to halt the process and
incorporate less Ag led to phase segregation, with Ag$_2E$ nucleated
at the NC surfaces \cite{Robinson2007}. (3) When individual Ag
impurities were eventually incorporated \cite{Ayash2012}, the NCs
exhibited optical and electronic properties suggesting a
transformation from a donor to an acceptor with increasing Ag
concentration.

In this Letter we propose an atomistic model that explains these
observations.  We used density-functional theory (DFT) to calculate
impurity formation energies and reaction barriers for Ag in bulk CdSe.
These were then used in dynamical simulations of cation exchange in
NCs at finite temperature over long times.  Our simulations
spanned a wide range of Ag concentrations, from light doping to complete
cation exchange, and provide a single conceptual framework in which to
understand the process. Our simulations show that Coulomb interactions
between ions play a central role in cation exchange. Indeed, if we
suppress Coulomb interactions then cation exchange does not occur.
Our simulations also show that Coulomb effects underlie the unusual
optical and electronic behavior of Ag-doped CdSe NCs.

We begin by considering two types of Ag impurities in bulk CdSe:
interstitial Ag (Ag$_{\mathrm{int}}$) and Ag substitutional on a Cd
site (Ag$_{\mathrm{Cd}}$) \cite{footnote1}. We used DFT total-energy
calculations to determine the equilibrium geometries and stable charge
states of each. Total energies and forces were calculated within the generalized-gradient
approximation of Perdew, Burke, and Ernzerhof (PBE) \cite{Perdew1996} using
projector-augmented-wave potentials, as implemented in {\sc vasp}
\cite{kresse_phys_rev_b_1993a,Kresse1996}.  The plane-wave cutoff was
300 eV.  We used an orthorhombic supercell with 360 atoms and sampled
the $\Gamma$ point.  We then used DFT with the hybrid functional of
Heyd, Scuseria, and Ernzerhof \cite{HSE06} to compare the stability of
different charge states. We find that Ag$_{\mathrm{int}}$ is a donor
with a binding energy of 0.10 eV, while Ag$_{\mathrm{Cd}}$ is an
acceptor with a binding energy of 0.32 eV. Hence the stable charge
states of Ag$_{\mathrm{int}}$ and Ag$_{\mathrm{Cd}}$ are $+$1 and
$-$1, respectively, when the Fermi level falls inside a window of
$\sim$1.2 eV within the band gap of 1.85 eV. We assumed these
conditions to hold throughout this study.  CdSe native defects could
in principle also be included in the model. However, we expect them to
be less important than the impurities themselves and therefore do not
include them here.

\begin{figure}[t]
\begin{center}
\includegraphics[width=0.75\textwidth]{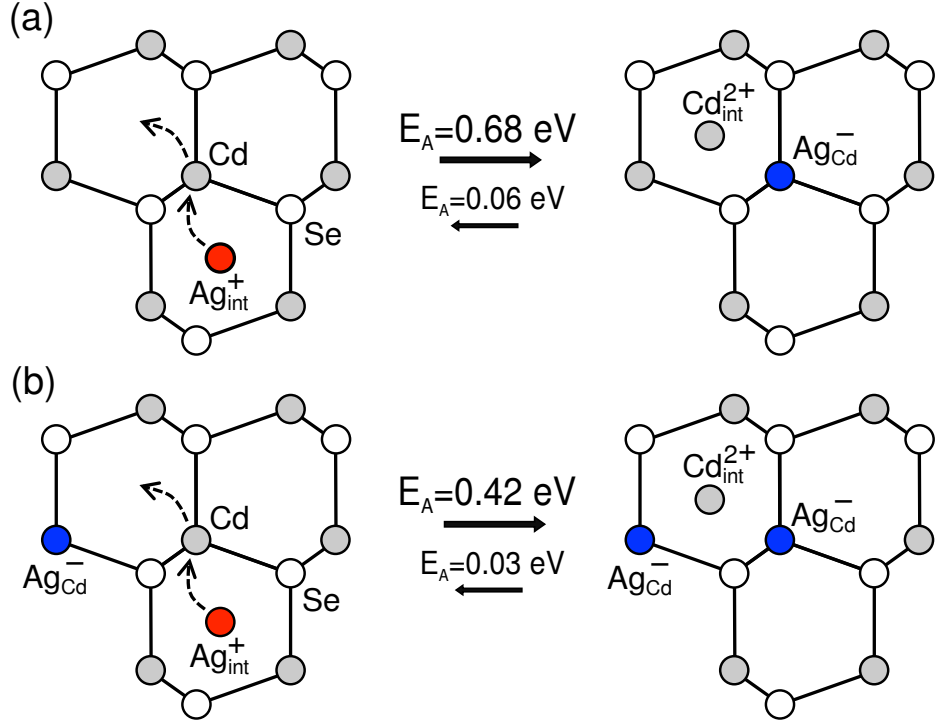}
\caption{Electrostatic reduction of the activation barrier for
    cation exchange.  For cation exchange to occur, Ag ions must
  replace Cd ions on the CdSe lattice.  (a) The pathway with the
  lowest activation barrier (0.68 eV in DFT) is the Cd kick-out
  reaction, in which an interstitial Ag$_{\rm int}^+$ displaces a Cd,
  creating an interstitial Cd$_{\rm int}^{2+}$ and a substitutional
  Ag$_{\rm Cd}^-$. (b) The barrier for the Cd kick-out reaction 
    is reduced (to 0.42 eV) if a substitutional Ag$_{\rm Cd}^-$
  already occupies a neighboring site, because the arrangement of
    charges in the final state is then electrostatically favorable.}
\label{fig1}
\end{center}
\end{figure}

We also used DFT with the PBE functional to calculate the activation barriers for two
important processes: diffusion of Ag$_{\mathrm{int}}$ and cation
substitution Ag$_{\mathrm{int}}$ $\rightarrow$ Ag$_{\mathrm{Cd}}$ $+$
Cd. The calculations used the nudged-elastic-band method in a 96-atom
supercell \cite{NEB}, and assumed the charge states discussed above
along with a compensating uniform background.  We find that
Ag$_{\mathrm{int}}$ moves between interstitial sites with a small
barrier of only 0.2 eV, consistent with the observed fast diffusion of
Ag in CdSe
\cite{Shaw1992, AgDiff}. 
It is reasonable to assume that, at room temperature, Ag ions enter
the NC from solution as Ag$_{\mathrm{int}}$ and then rapidly diffuse
between interstitial sites \cite{footnote4}.  We
investigated several reaction pathways for cation substitution and
found that the kick-out reaction [Fig.\ \ref{fig1}(a)] had by far the
lowest activation barrier, 0.68 eV \cite{footnote2}. In this reaction,
Ag$_{\mathrm{int}}$ moves onto a Cd site by displacing the Cd to an
interstitial site. As a result, Ag$_{\mathrm{Cd}}$ and an interstitial
Cd (Cd$_{\mathrm{int}}$) are created. The stable charge state of
Cd$_{\mathrm{int}}$ is $+2$ (donor binding energy 0.29 eV) and thus
the reaction preserves total charge: Ag$_{\mathrm{int}}^+$ $+$
Cd$_{\mathrm{Cd}}^0$ $\rightarrow$ Ag$_{\mathrm{Cd}}^-$ $+$
Cd$_{\mathrm{int}}^{2+}$.

The final state of this reaction is only marginally
stable because the reverse reaction has a very low barrier (0.06
eV). Consequently, after a kick-out step the system usually reverts to
its initial state.  Evidently, the cation-exchange process can 
proceed only if the Ag-Cd complex dissociates.  This can occur 
if the Cd diffuses away, leaving behind Ag$_{\mathrm{Cd}}$.
Because the rate of this dissociation governs the overall
cation-exchange rate, it is important to evaluate it under realistic
conditions---that is, in small NCs containing many charged Ag impurities.

One consequence of confining ionized impurities to the small volume of
a NC is that strong Coulomb interactions are inevitable. These
profoundly affect the spatial distribution and reaction barriers of
the impurities, and must therefore be included in a microscopic
model. For example, if a NC contains several Ag$_{\mathrm{int}}$ with
the same charge, then their mutual repulsion pushes them toward the
NC surface.  Each ion is also influenced by the polarization at
the dielectric interface between the NC and the surrounding medium.
Because the semiconductor dielectric constant is typically larger than
its surroundings, this interaction is always repulsive and hence
pushes the impurities toward the NC interior. The outcome of these
competing interactions becomes more complicated when the dopants are
of different types, are in different charge states, or occupy both
interstitial and substitutional sites.

The importance of Coulomb effects is especially clear
for Cd kick-out reactions. In the absence of other dopants,
this reaction is energetically uphill with a substantial activation
barrier (0.68 eV). For an attempt frequency
of $10^{13}$ Hz \cite{Voter2007}, this barrier can be overcome at
room temperature, but not at a rate consistent with 
the observed millisecond time scales \cite{Chan2007a}. Experiments suggest much
higher rates and thus lower activation barriers. This
puzzle is resolved by considering Coulomb interactions.  If
other ionized impurities are nearby, they can strongly alter the barrier
for reactions in their vicinity.
Indeed, we find using DFT that Coulomb interactions greatly reduce
the barrier for kick out whenever the final state
is favored by the arrangement of charges.  In the best case, a
nearby Ag$_{\mathrm{Cd}}$ lowers the barrier from
0.68 to 0.42 eV [Fig.\ \ref{fig1}(b)], because in the final state the
Cd$_{\mathrm{int}}^{\mathrm{2+}}$ is located between oppositely charged
(Ag$_{\mathrm{Cd}}^{\mathrm{-}}$) impurities. At room temperature this
reduction increases the kick-out rate by a factor
of 10$^4$.

Coulomb interactions also drive the cation-exchange process
on a larger scale, by favoring the formation of contiguous regions of
electrostatically bound Ag$_2$ pairs (Ag$_{\mathrm{int}}^+$ plus
Ag$_{\mathrm{Cd}}^-$) for each Se atom. {\em Thus, the
  cation-exchange reaction depends on a positive feedback arising from
  strong Coulomb interactions between charged impurities.}  This
conclusion agrees with experiments showing that cation
exchange in NCs is a cooperative mechanism in which one
event triggers an extremely fast transformation
\cite{White2013}.  In our case, this event is the formation of the
initial Ag$_{\mathrm{Cd}}$, which we hypothesize occurs by Cd kick out
at the NC surface.  We also found using
  DFT \cite{footnote1} that such a fully substituted 
phase is crystallographically---as well as energetically---very close to the
crystalline Ag$_2$Se naumannite phase, consistent with experiment
\cite{Robinson2007}.

\begin{figure}[tl]
\begin{center}
\includegraphics[width=0.75\textwidth]{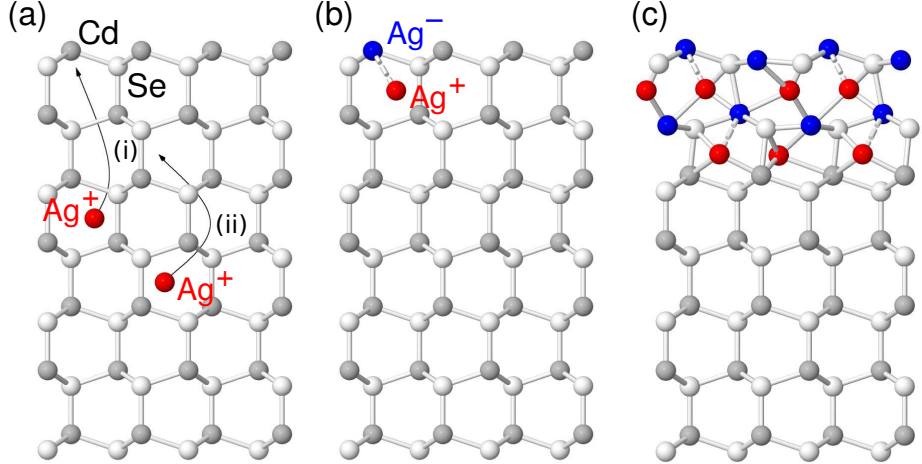}
\caption{Microscopic model of Ag doping and cation exchange of CdSe
  NCs.  (a) At low concentration, Ag are highly mobile interstitial
  ions, Ag$_{\mathrm{int}}^{\mathrm{+}}$. As their number increases,
  their mutual repulsion pushes them toward the NC surface (top) where
  a Cd kick-out reaction can occur [step (i)]. The substitutional
  Ag$_{\mathrm{Cd}}^{\mathrm{-}}$ created by the kick-out reaction
  attracts the remaining Ag$_{\mathrm{int}}^{\mathrm{+}}$ [step (ii)].
  (b) An electrostatically bound complex is then formed by two Ag at
  the surface, forming a nucleus of Ag$_2$Se. (c) At even
  higher Ag concentrations, all interstitial and Cd sites at the surface become
  occupied with Ag. This fully substituted CdSe closely resembles the
  naumannite crystal structure of Ag$_2$Se.}
\label{fig2}
\end{center}
\end{figure}

Figure \ref{fig2} summarizes the steps in our model.  Ag ions enter
the NC via interstitial pathways. Repelled from the surface by
dielectric polarization, they diffuse towards the center. As the
concentration increases, their mutual repulsion becomes dominant and
drives them back towards the surface [Fig.\ 2(a)]. At this stage, the
formation of Ag$_{\mathrm{Cd}}$ is likely to occur, for two
reasons. (1) The formation of negatively charged Ag$_{\mathrm{Cd}}^-$
can compensate electrically for the positive charge accumulating near
the surface due to the Ag$_{\mathrm{int}}^+$. (2) The Cd that is replaced by
Ag can immediately leave the NC by diffusing into the surrounding
solution, preempting the reverse kick-out reaction. Once
Ag$_{\mathrm{Cd}}^{\mathrm{-}}$ is formed, it attracts mobile
Ag$_{\mathrm{int}}^{\mathrm{+}}$ [Fig.\ \ref{fig2}(b)] and, in
addition, lowers the barrier for nearby kick-out reactions. 
Consequently, additional interstitial and substitutional Ag coalesces
in the surface region around the nucleating site
[Fig.\ \ref{fig2}(c)]. The result is a stable and densely
packed cluster of alternating charges, having the stoichiometry of
Ag$_2$Se and an arrangement of atoms close to the naumannite
structure, into which it easily relaxes.

To investigate our model quantitatively we used dynamical simulations
based on the kinetic Monte Carlo (KMC) method \cite{Voter2007}, an
efficient technique for simulating dynamics
over long times at finite temperature.  An initial random state
was stochastically evolved by selecting from a list of
processes (diffusion and kick-out reactions) using
acceptance ratios given by their Boltzmann factors. These were
evaluated using the DFT barriers already computed
\cite{Vineyard1957}. To account for Coulomb interactions, we added the state-dependent Coulomb
potential from all ions to the reaction pathway for each
process \cite{footnote3}.  When the impurities were within approximately one lattice
constant [as in Fig.\ 1(a)], these were included
directly using DFT. For larger separations they were
approximated using point charges and the CdSe dielectric constant.
Interactions with polarization charges at the nanocrystal surface were
included using an electrostatic model \cite{Batygin1965} 
of point charges in a
dielectric sphere embedded in an organic medium.

\begin{figure}[t]
\begin{center}
\includegraphics[width=0.75\textwidth]{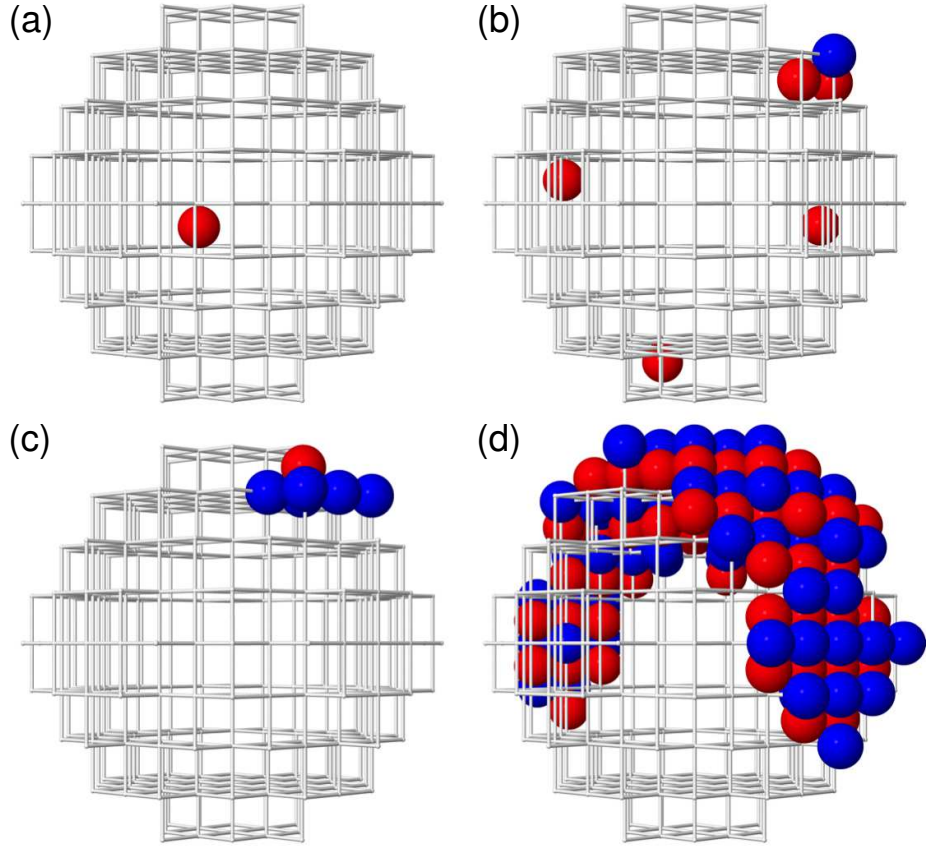}
\caption{Four stages of cation exchange, depicted by snapshots
  from KMC simulations at 300 K. Red and blue spheres denote
  Ag$_{\mathrm{int}}^+$ and Ag$_{\mathrm{Cd}}^-$, respectively. (a)
  With only one Ag$_{\mathrm{int}}^+$, the impurity stays near the NC
  center. (b) With additional Ag$_{\mathrm{int}}^+$, they repel each
  other toward the surface, where kick out of Cd by Ag can occur. (c)
  If no further Ag is added, the impurities eventually
  cluster at the surface, mostly as Ag$_{\mathrm{Cd}}^-$. (d) Additional
  Ag leads to the growth of Ag$_2$Se at the surface.}
\label{fig3}
\end{center}
\end{figure}

We performed the simulations for a quasispherical NC with a
realistic diameter of 3.5 nm [Fig.\ \ref{fig3}]. For 
simplicity we used a simple cubic lattice and set
the density of Cd sites to that of wurtzite CdSe, giving a
lattice constant of 3.9 \AA. The Se atoms were included implicitly and
considered immobile because they are known not to participate in
cation exchange.  Ag ions were added to
the NC at interstitial sites on the surface. In the absence of any
Coulomb interactions, the reaction barrier for this step was set to
0.6 eV to yield reasonable addition rates for Ag.  We also assumed
that Ag$_{\mathrm{int}}$ did not leave the NC.  In
contrast, Cd$_{\mathrm{int}}$ was allowed to diffuse out,
as is observed experimentally \cite{Son2004}; the driving force
for this is the hard Lewis base (e.g.~methanol or ethanol)
that comprises the solution and which strongly binds to the hard Lewis
acid Cd$^{2+}$. In the real system, Cd ions hence become solvated
in solution; for simplicity we removed them from
the simulation after they diffused out.

As Ag$^{+1}$ ions enter and Cd$^{+2}$ ions leave, the NC may become
transiently charged. This charge must be balanced by
counterions outside the NC. We assumed these to be mobile and thus
treated them as uniformly spread out near the NC; hence
they did not affect the potential-energy surface for reactions inside 
\cite{footnote1}. On average, two Ag$^{+1}$ ions enter for each
Cd$^{+2}$ that leaves, and therefore the NC maintains approximate
electrical neutrality as CdSe is converted to Ag$_2$Se.

Figure \ref{fig3} shows snapshots from simulations at 300 K. Four
different stages are depicted, from one Ag to a partially
cation-exchanged NC with Ag$_2$Se formed near the surface.  The
simulations confirm that, due to the dielectric polarization discussed
above, a single Ag$_{\mathrm{int}}^{\mathrm{+}}$ tends to stay in the
NC center as a stable interstitial [Fig.\ \ref{fig3}(a)], whereas
several Ag$_{\mathrm{int}}^{\mathrm{+}}$ push each other toward
the surface and allow the formation of Ag$_{\mathrm{Cd}}$ [Fig.\
\ref{fig3}(b)]. If the supply of Ag$_{\mathrm{int}}$ is stopped at
this stage, the simulation  evolves to a configuration with a
majority of Ag$_{\mathrm{Cd}}$ [Fig.\ \ref{fig3}(c)]. If further Ag is
then provided, a well-defined Ag$_2$Se phase grows from
the surface inwards [Fig.\ \ref{fig3}(d)]. Although we did not run the
simulations to completion (due to their cubic dependence on the number
of Ag ions) it is clear that, if continued, the nanocrystal would
have been completely converted to Ag$_2$Se.  A video animation of one
simulation run is available in the Supplemental Material \onlinecite{footnote1}.

\begin{figure}[t]
\begin{center}
\includegraphics[width=0.75\textwidth]{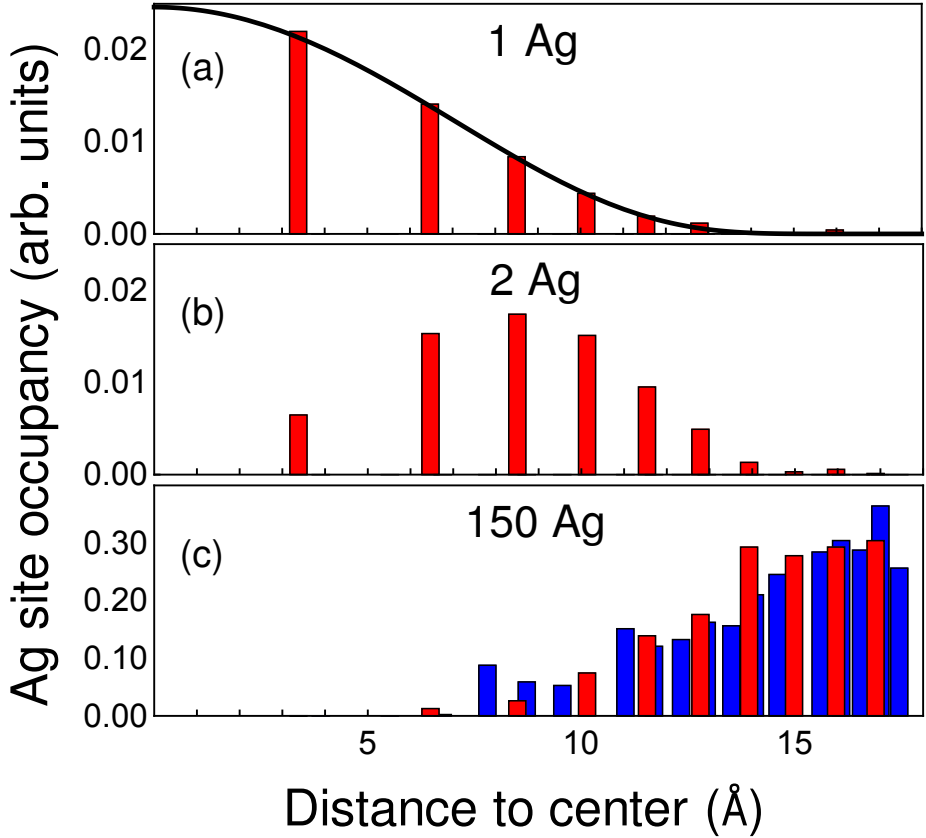}
\caption{Site occupancy at 300 K of Ag$_{\mathrm{int}}$ (red) and
  Ag$_{\mathrm{Cd}}$ (blue) versus radial distance from the center of
  the NC. The distributions are averages over 100 KMC
  simulations for (a) one, (b) two, and (c) 150 Ag impurities per NC. The
  black curve in (a) shows the Boltzmann distribution at 300 K
based on the energy from dielectric polarization.}
\label{fig4}
\end{center}
\end{figure}

These simulations can be investigated more quantitatively by
averaging over many runs.  Figure \ref{fig4}
shows the Ag site occupancy as a function of radial position for NCs
with one, two, and 150 Ag dopants. The results
reflect the competition between opposing Coulomb effects.
With a single Ag impurity, interstitial sites near the 
  center are strongly favored, because dielectric polarization
 dominates the interactions. This is clear from
comparing to the distribution predicted analytically [black curve in Fig.\
\ref{fig4}(a)].  With only one impurity, the formation of
Ag$_{\mathrm{Cd}}$ is unlikely because Ag$_{\mathrm{int}}$, despite
being mobile, spends very little time near the surface.  However, even
with only two Ag$_{\mathrm{int}}^{\mathrm{+}}$, the ion-ion
interactions change the shape of the concentration
distribution [Fig.\ \ref{fig4}(b)].  With 150 Ag, both
Ag$_{\mathrm{int}}$ and Ag$_{\mathrm{Cd}}$ are present and their
concentration is greatest at the surface [Fig.\ \ref{fig4}(c)].  The
two types of Ag are nearly equal due to the formation of Ag$_2$Se, because
naumannite has two inequivalent Ag sites.

Our model is also consistent with two trends observed
experimentally in Ag-doped CdSe NCs \cite{Ayash2012}.  (1) The
photoluminescence (PL) efficiency of undoped samples 
increased dramatically even when only one Ag was added per NC.  Our
simulations predict that a single Ag$_{\mathrm{int}}^{\mathrm{+}}$
preferentially occupies the NC center.  This positive charge
can influence the photoexcited exciton by pulling the electron away
from the surface and suppressing nonradiative
recombination.  Moreover, the PL efficiency  decreased
with further Ag incorporation, consistent with the predicted
redistribution of Ag toward the surface. (2) Electrical transport
measurements on doped NC films showed an increase and then
a decrease in the Fermi level when the number of Ag per NC was varied
from zero to 20. This was attributed to a change of the Ag from
interstitial donors to substitutional acceptors at $\sim$7 Ag per NC
\cite{Ayash2012}.  Our results confirm that at low concentration, Ag
persists as an interstitial donor and is easily ionized to provide
free electrons.  At higher Ag concentrations the majority become
substitutional acceptors [Fig.\ 3(c)]. Thus, our simulations are
consistent with the observed nonmonotonic dependence of the Fermi
level on Ag incorporation.  Finally, at higher concentrations,
Ag$_2$Se starts to form at the surface, and the behavior of the Fermi
level becomes more complicated.
 
In summary, we have proposed a microscopic model of Ag cation exchange
in CdSe NCs.  The key mechanisms---interstitial diffusion, cation
kick out, impurity-impurity Coulomb interactions, and NC surface
polarization---are quite general and therefore likely relevant to
other cation-exchange systems, such as Cu in Cd$E$. Furthermore, because
more complex situations with mixed electrically active dopants can now
be treated, we anticipate that new and unexpected phenomena will arise
when these principles are applied to other NC systems.

\begin{acknowledgments}
  We thank Al.~L.~Efros, A.~Sahu, and V.~Holmberg for helpful
  discussions.  This work was supported by the Swiss National Science
  Foundation under Award no. 200021-140617 and by the U.S. Office of
  Naval Research through the Naval Research Laboratory's Basic
  Research Program (SCE). Computations were performed at the ETH
  High-Performance Computing Cluster Brutus and the DoD Major Shared
  Resource Center at AFRL.
\end{acknowledgments}


\begin{thebibliography}{27}
\bibitem[{foo({\natexlab{b}})}]{footnote2}
\bibinfo{note}{Other mechanisms, such as concerted exchange and
  vacancy-mediated exchange, have higher barriers (2.0 and 1.0 eV,
  respectively).}
\bibitem[{foo({\natexlab{c}})}]{footnote3}
\bibinfo{note}{For simplicity we evaluated the Coulombic contribution to the
  activation barrier by assuming the transition state to be midway between the
  initial and final states.}
\end{thebibliography}
\end{document}